\documentclass[10pt,lettersize,journal]{IEEEtran}

\usepackage{amsmath,amsfonts}
\usepackage{algorithmic}
\usepackage{algorithm}
\usepackage{array}
\usepackage{textcomp}
\usepackage{stfloats}
\usepackage{verbatim}
\usepackage{graphicx}
\usepackage{cite}
\usepackage[table]{xcolor}
\usepackage{multirow}
\usepackage{subcaption} 
\usepackage{float}
\usepackage{placeins}
\usepackage{wrapfig}
\usepackage[hidelinks,colorlinks=true,linkcolor=black,citecolor=black,urlcolor=black]{hyperref}
\usepackage{makecell}
\usepackage{tikz}
\usepackage{enumitem}
\usepackage{titlesec}
\titlespacing*{\section}{0pt}{*0.8}{*0.8}
\titlespacing*{\subsection}{0pt}{*0.6}{*0.6}
\titlespacing*{\subsubsection}{0pt}{*0.5}{*0.5}

\begin{document}

\title{Reducing Communication Overhead in Federated Learning for Network Anomaly Detection with Adaptive Client Selection}

\author{ William Marfo\textsuperscript{\textdagger}, Deepak K. Tosh\textsuperscript{\textdagger}, Shirley V. Moore\textsuperscript{\textdagger}, Joshua Suetterlein\textsuperscript{*}, Joseph Manzano\textsuperscript{*}\\
\textsuperscript{\textdagger}Department of Computer Science, \textsuperscript{*}High-Performance Computing Group\\
\textsuperscript{\textdagger}University of Texas at El Paso, El Paso, TX 79968, USA\\
\textsuperscript{*}Pacific Northwest National Laboratory, Richland, WA 99354, USA\\
wmarfo@miners.utep.edu, \{dktosh, svmoore\}@utep.edu, \{joshua.suetterlein, joseph.manzano\}@pnnl.gov }



\maketitle

\begin{abstract}
Communication overhead in federated learning (FL) poses a significant challenge for network anomaly detection systems, where the myriad of client configurations and network conditions can severely impact system efficiency and detection accuracy. While existing approaches attempt to address this through individual optimization techniques, they often fail to maintain the delicate balance between reduced overhead and detection performance. This paper presents an adaptive FL framework that dynamically combines batch size optimization, client selection, and asynchronous updates to achieve efficient anomaly detection. Through extensive profiling and experimental analysis on two distinct datasets—UNSW-NB15 for general network traffic and ROAD for automotive networks—our framework reduces communication overhead by \textit{97.6\%} (from \textit{700.0s} to \textit{16.8s}) compared to synchronous baseline approaches while maintaining comparable detection accuracy (\textit{95.10\%} vs. \textit{95.12\%}). Statistical validation using Mann-Whitney \(\mathrm{U}\) test confirms significant improvements $(p < 0.05)$ over existing FL approaches across both datasets, demonstrating the framework's adaptability to different network security contexts. Detailed profiling analysis reveals the efficiency gains through dramatic reductions in GPU operations and memory transfers while maintaining robust detection performance under varying client conditions.
\end{abstract}
\section{Introduction}
\label{sec:introduction}

The rapid growth of digital connectivity and network infrastructure, driven by the proliferation of connected devices and critical digital systems, has significantly increased the potential for network anomalies, including cyber-attacks and system faults \cite{marfo2022network}. Detecting these anomalies swiftly and accurately is crucial for maintaining the security and reliability of networked environments, particularly in sectors where uninterrupted service is essential \cite{nishio2019client}. While network anomaly detection aims to identify deviations in traffic behavior that may indicate threats, traditional centralized machine learning (ML) methods, which aggregate data on central servers, often face issues with latency, scalability, and data transfer costs, making them unsuitable for large-scale network anomaly detection \cite{marfo2024enhancing}.

Federated learning (FL) has emerged as a viable alternative, enabling decentralized model training across distributed client nodes. FL allows data to remain localized while model updates are communicated to a central server, reducing the need for centralized data collection \cite{lee2024fedl2p}. This decentralized approach makes FL particularly suited for time-sensitive applications, such as network anomaly detection, where timely responses to evolving patterns are critical. Despite these advantages, FL introduces a new challenge: managing the communication overhead associated with frequent model updates, especially in time-sensitive applications like network anomaly detection, where delays can compromise real-time detection capabilities \cite{luping2019cmfl}. Efficiently balancing communication overhead with model performance remains a key barrier to the effective deployment of FL for network anomaly detection.

Existing approaches to reduce FL communication overhead include techniques like client clustering and selective updates \cite{al2021reducing}. While these methods provide useful insights, they often address isolated aspects of the problem and lack a comprehensive framework that dynamically balances communication efficiency with detection performance. This limitation is further compounded by the diverse conditions across clients, such as variations in computational resources and network quality, which can disrupt system robustness and consistency \cite{marfo2024enhancing,marfo2022network}.

In response to these challenges, we propose an adaptive FL framework specifically designed to optimize communication overhead in network anomaly detection. Our framework incorporates the following contributions:
\begin{itemize}
    \item A dynamic batch size optimization strategy coupled with efficient local training through mixed precision and distributed data parallelism (DDP) to enhance computational efficiency on client devices.
    \item An asynchronous communication approach to minimize idle time and reduce bottlenecks, improving overall communication efficiency.
    \item A gradient alignment-based selective update mechanism that filters out unconstructive updates, ensuring only the most relevant updates contribute to the global model.
\end{itemize}

We validate our framework on two datasets: UNSW-NB15 \cite{nour2016unsw}, representing modern network traffic, and ROAD \cite{verma2022addressing}, containing automotive network data with verified attacks. In automotive networks, detecting anomalies like masquerade attacks is critical, as they can manipulate vehicle controls, causing unintended braking or sudden acceleration, posing serious safety risks. Through detailed profiling and experimental analysis, our results demonstrate substantial reductions in communication overhead and enhanced computational efficiency across both datasets, establishing a robust foundation for deploying FL in various network security contexts. To advance the state of the art in FL for network anomaly detection, we make the code for our framework publicly available on GitHub~\cite{Nsight-Profiling-Visualization}. This enables other researchers to replicate our results, compare new algorithms, and further explore communication-efficient FL techniques.

The remainder of this paper is organized as follows: Section \ref{sec:background} outlines key FL challenges. Section \ref{sec:relatedwork} reviews related work on communication optimization in FL. Section \ref{sec:framework} describes our proposed optimization methods. Section \ref{sec:results} presents experimental setup and results, followed by a discussion in Section \ref{sec:Discussion}. Finally, Section \ref{sec:conclusion} summarizes the findings and suggests future research directions.

\section{Background} \label{sec:background}

\subsection{Communication Overhead and Synchronous Updates in FL}
\label{subsec:comm_overhead}

Synchronous FL algorithms, such as FedAvg \cite{mcmahan2017communication}, inherently face significant challenges when deployed in real-world environments. The core issue arises from the ``straggler effect," where a central server must wait for all participating clients to complete their local training before model aggregation \cite{mcmahan2017communication}. This synchronization requirement means the entire system operates at the speed of the slowest client, creating a fundamental bottleneck in the training process. The impact of these delays is particularly problematic in network anomaly detection scenarios, which demand timely responses to security threats. Delays in aggregating model updates can result in postponed anomaly detection, allowing security threats to persist for extended periods before being addressed \cite{al2021reducing}. Traditional solutions that attempt to address this by simply excluding slow clients are problematic, as they can lead to bias in the trained model and potentially miss important anomaly patterns present in the data of slower clients \cite{nishio2019client}.

\subsection{Batch Size Considerations in FL}
\label{subsec:batch_size}
Batch size in FL presents unique challenges compared to centralized training due to decentralized devices with varying computational capabilities and data distributions \cite{al2021reducing}. In FL, a \textit{communication round}—where clients perform local training, transmit model updates to the server, and receive an aggregated global model—creates fundamental trade-offs. Larger batch sizes reduce communication rounds \cite{luping2019cmfl} but can impair generalization and increase local computation time, particularly with heterogeneous devices and non-IID data \cite{marfo2022network}. While smaller batch sizes enhance generalization and convergence, they increase communication overhead \cite{lee2024fedl2p}. This trade-off becomes critical in anomaly detection scenarios, where non-IID data distributions can amplify update variance with larger batches \cite{yan2023criticalfl}, while smaller batches, despite enabling faster initial convergence, incur higher network communication costs \cite{luping2019cmfl}.

\section{Related Work} \label{sec:relatedwork}

Al-Saedi et al. \cite{al2021reducing} proposed CA-FL to reduce communication overhead in FL by clustering worker updates based on similarity. Their analysis of human activity recognition (HAR) datasets showed that clustering similar updates effectively reduces communication costs and maintains model accuracy. This makes CA-FL a promising solution for improving efficiency without compromising performance. 

Hu et al. \cite{hu2022dpro} introduced dPRO, a performance diagnosis toolkit for distributed deep neural network (DNN) training. Their approach, tested across TensorFlow and MXNet frameworks, identified performance bottlenecks in computation, communication, and memory. This resulted in up to a 3.48× speedup over baseline methods, demonstrating dPRO's ability to significantly enhance distributed training performance through targeted optimizations.

Miao et al. \cite{miao2022md} presented MD-Roofline, an extension of the traditional roofline model that incorporates communication dimensions to analyze distributed deep learning training performance. Using 12 classic convolutional neural networks (CNNs), they demonstrated how MD-Roofline could pinpoint bottlenecks across intra-GPU computation capacity, memory access bandwidth, and inter-GPU communication, providing valuable insights for optimizing DNN training.

Aach et al. \cite{aach2023large} performed a large-scale evaluation of distributed deep learning frameworks, comparing tools like Horovod and DeepSpeed in terms of runtime performance and scalability. They trained ResNet architectures on the ImageNet dataset using up to 1024 GPUs and demonstrated that optimizing the choice of data loaders and frameworks could dramatically reduce training time from 13 hours to just 200 seconds, highlighting the impact of distributed deep learning on accelerating model development.

Wen et al. \cite{wen2023survey} conducted a comprehensive survey on FL, focusing on the various challenges and applications. They specifically identified communication overhead as a critical bottleneck in FL that requires innovative solutions, underscoring the need for further research into efficient communication strategies.

Li et al. \cite{li2024understanding} conducted a systematic analysis of communication characteristics in distributed training environments. They developed analytical formulations to estimate communication overhead, considering various influencing factors. Their work emphasizes the need for a deep understanding of communication patterns to inform optimizations for distributed training systems.

Previous works have advanced understanding of communication overhead in federated and distributed learning but often focus on isolated aspects such as clustering, performance analysis, or framework optimization. A critical gap remains in balancing communication efficiency and model performance, particularly for time-sensitive tasks like network anomaly detection. We address this gap with three key innovations: (1) an adaptive FL framework using batch size optimization, mixed precision, and DDP for efficient local training, (2) an asynchronous communication strategy tailored for FL, which reduces idle time and improves scalability in heterogeneous environments, and (3) a gradient alignment-based selective update mechanism to integrate only constructive updates. Unlike traditional asynchronous methods used in parallel DNN training, our approach is specifically designed for FL, addressing challenges such as stragglers, non-IID data, and varying client conditions. Our validation on the UNSW-NB15 \cite{nour2016unsw} and ROAD \cite{verma2022addressing} datasets demonstrates substantial improvements in both communication efficiency and detection accuracy.

\section{Proposed FL Optimization Methods for Network Anomaly Detection}
\label{sec:framework}

\begin{figure*}[htbp]
    \centering
    \includegraphics[width=0.98\textwidth]{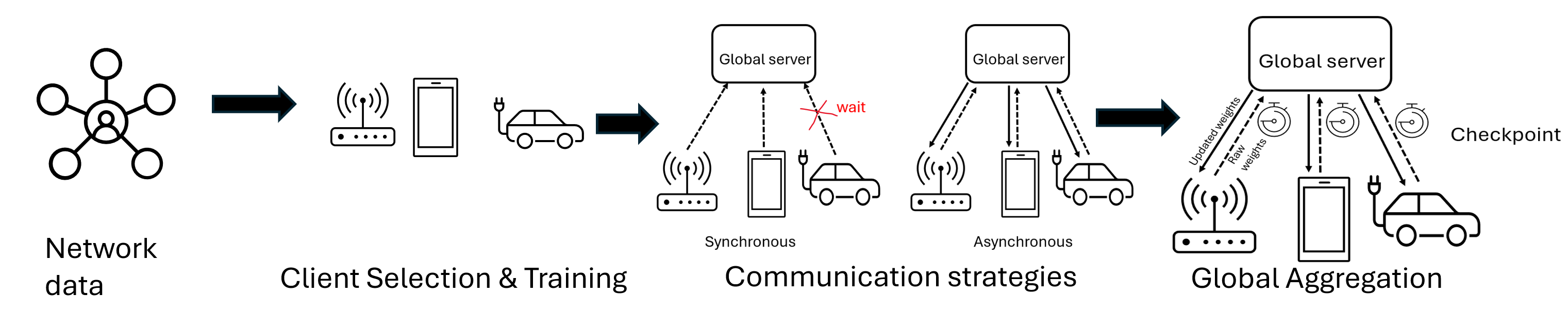}
    \caption{Framework architecture integrating efficient client training, asynchronous communication, and selective updates.}
    \label{fig:workflow}
\end{figure*}

This section introduces our proposed optimization methods for improving FL in network anomaly detection. Our framework addresses three key challenges in distributed anomaly detection through integrated solutions: (1) reducing communication overhead via selective updates, (2) improving computational efficiency through efficient local training, and (3) mitigating accuracy degradation using asynchronous communication. Fig.~\ref{fig:workflow} illustrates the flow of data through our framework, from initial client-side processing to global model aggregation.

\subsection{Efficient Local Training}
\label{subsec:local_training}

Our FL system is composed of two main components: \textit{clients} and a \textit{global server}. Each client \( c_i \) trains a local model \( f_i \) on its dataset \( X_i \), with model parameters \( w_i \). After local training for \( e \) epochs, each client transmits its updated parameters \( w_i \) to the global server. The server aggregates these parameters using an averaging process to update the global model \( w_g = \frac{1}{N} \sum_{i=1}^{N} w_i \), where \( N \) is the number of participating clients.

We combine three key optimizations to reduce computational overhead while maintaining detection accuracy. Batch size significantly impacts training efficiency in distributed environments \cite{aach2023large}. Given the varying data characteristics and distributions across our datasets, larger batch sizes enable more efficient data processing, which is crucial for minimizing communication rounds while preserving the ability to capture subtle variations indicating anomalies. Larger batch sizes reduce communication frequency between clients and the global server by requiring fewer gradient updates, though they require careful tuning to avoid compromising model convergence. Our framework dynamically adjusts batch sizes based on client capacity, balancing communication overhead against convergence requirements.

\textbf{Dynamic batch size adjustment:}
During training, each client reports local metrics (GPU utilization, memory usage, network latency) to the server, which assigns a batch size proportional to the client’s available resources. For example, a high-capacity client might train with 512 samples per batch to reduce communication rounds, whereas a lower-capacity client uses 64 to prevent straggler delays and maintain convergence quality. This mechanism ensures that each client operates at an optimal pace, balancing communication overhead with accurate model updates.

We implement mixed precision training using \texttt{autocast} and \texttt{GradScaler}, which reduces memory usage and accelerates computation by leveraging \texttt{float16} for most operations while retaining \texttt{float32} for critical components \cite{nvidia_amp}. This approach is beneficial for high-dimensional anomaly detection: \texttt{float16} operations execute faster on modern GPUs, while \texttt{float32} prevents gradient underflow and preserves numerical stability. During local training, \texttt{autocast} automatically casts forward passes to \texttt{float16}, and \texttt{GradScaler} manages gradient scaling to avoid vanishing gradients. As a result, clients experience lower computational overhead without compromising detection accuracy.

To efficiently utilize available computational resources, we employ DDP training \cite{aach2023large}, where each client trains the local model across multiple GPUs with synchronized gradient updates. Our implementation runs in a high-performance computing environment using SLURM for resource management, configured with 4 GPUs per node and dedicated task resources (\texttt{--ntasks=4}, \texttt{--cpus-per-task=4}). This configuration enables efficient scaling of the training process while maintaining synchronization across distributed components.

\subsection{Communication Optimization}
\label{subsec:communication}

\begin{figure}[h]
\centering
\includegraphics[width=0.50\linewidth]{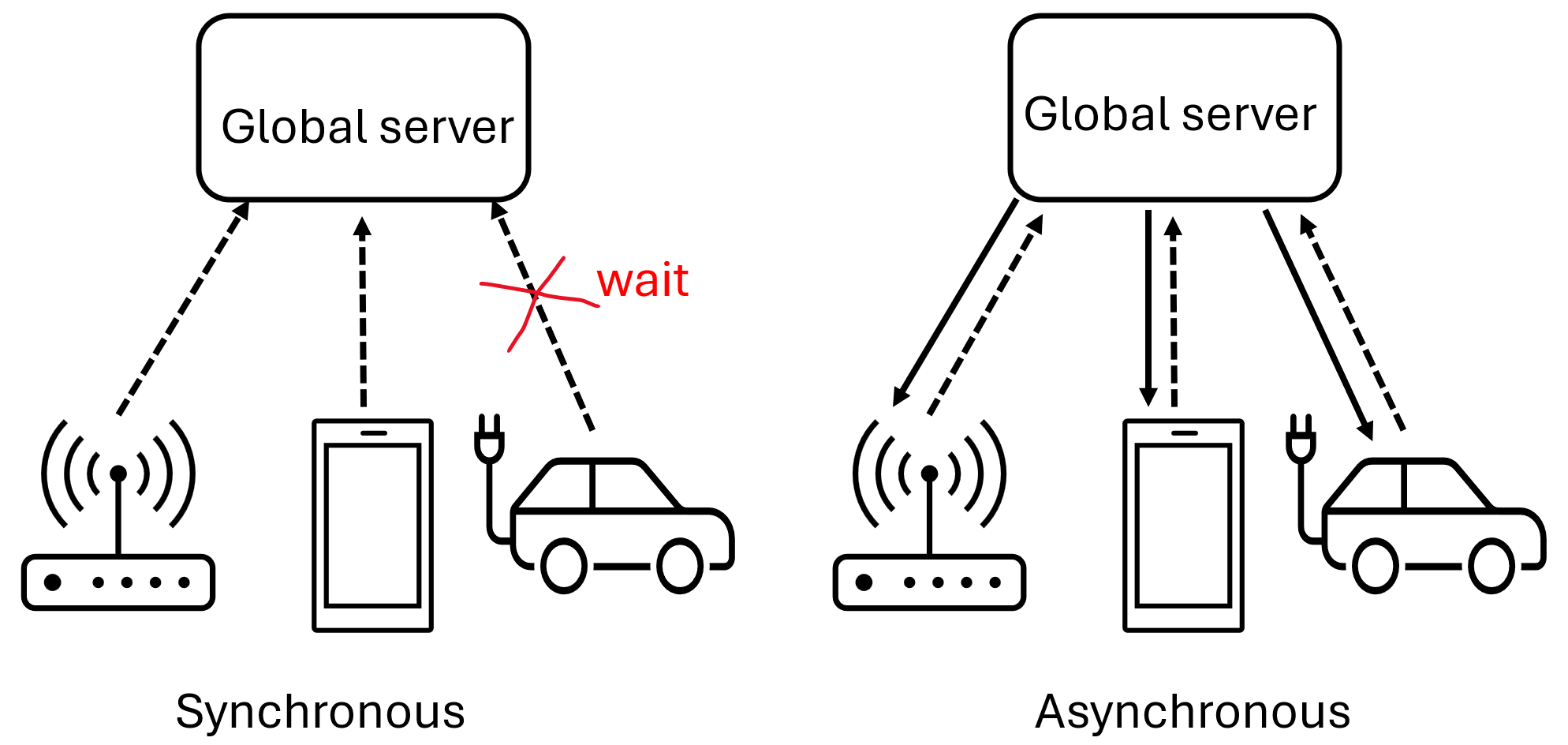}
\caption{Impact of synchronous (left) versus asynchronous (right) communication in FL. In the synchronous approach, the red “X” indicates the synchronization barrier, where the server waits for all clients to finish training before aggregation. Asynchronous updates allow continuous model improvement through independent client contributions without such a barrier.}
\label{fig:sync_vs_async}
\end{figure}

Traditional synchronous FL faces significant performance bottlenecks when clients operate at different speeds or experience varying network conditions \cite{xu2023asynchronous}. In synchronous updates, the global server must wait for all clients to complete their local training before model aggregation, resulting in increased idle times and communication overhead, especially when slower clients are involved \cite{luping2019cmfl}. Fig.~\ref{fig:sync_vs_async} illustrates this challenge, showing how system efficiency can be affected by client or network delays.

Our framework implements asynchronous communication, allowing concurrent client updates without global synchronization. This design reduces idle time by letting faster clients contribute updates immediately and improves fault tolerance through independent client operation. Clients train independently, submit updates once ready, and a thread pool manages concurrent submissions. The server aggregates updates continuously, eliminating delays from stragglers. Removing synchronization barriers enables continuous operation in heterogeneous environments with varying client capabilities and network conditions \cite{luping2019cmfl}.

\subsection{Global Model Aggregation}
\label{subsec:aggregation}

Our framework implements client-side filtering for efficient global model aggregation in FL, reducing communication overhead by filtering updates at the source based on gradient sign alignment. Specifically, after local training, each client compares the signs of local gradients with the last known global model gradients, calculating an alignment ratio as the proportion of parameters with matching signs. We set an alignment threshold of 65\% as the selection criterion through empirical search across a range of values, balancing the trade-off between filtering noisy updates and retaining constructive client contributions. Only clients achieving alignment ratios above this threshold transmit their model parameters \(w_i\) to the global server. The server then aggregates these pre-filtered updates into the global model as \(w_g = \frac{1}{|S|} \sum_{i \in S} w_i\), where \(S\) represents the set of clients that met the threshold criteria. The updated global model \(w_g\) is then broadcast to all clients to synchronize their models for the next training round. This client-side selective update mechanism integrates seamlessly with our asynchronous framework~\cite{aach2023large}, benefiting heterogeneous FL scenarios with diverse client data.

To enhance the fault tolerance of our system, we incorporate an adaptive checkpointing mechanism. Traditional FL systems often face challenges with node failures, typically resulting in complete restarts or reinitialization with the last known global weights, which leads to significant training disruptions. To address these challenges, we employ a dynamic checkpointing mechanism based on node failure probability modeling, thus enhancing system robustness. We use a Weibull distribution, represented as \( F(t) = 1 - \exp\left(-\left(\frac{t}{\lambda}\right)^k\right) \), where \( \lambda \) and \( k \) are parameters derived from historical failure data, to predict node failures. The optimal checkpointing interval \( t_c^* \) is determined by balancing overhead cost and recovery time through the cost function \( C(t_c) = \frac{t_c}{T} + p_f(t_c) \cdot \frac{t_r}{T} \), where \( T \) is the total computation time, \( t_r \) is the recovery time, and \( p_f(t_c) \) represents the failure probability at interval \( t_c \). Minimizing this function ensures that checkpoint intervals are efficient and adaptable to varying system conditions. During operation, each client maintains state information, including model parameters, optimization states, and training progress. If a client drops out, the system uses checkpoints to restore its state and resynchronize with the global model, avoiding full retraining.

By combining client-side filtering with asynchronous aggregation and adaptive checkpointing, our approach significantly reduces communication overhead while maintaining system robustness. This ensures that only constructive updates are transmitted, preserving training continuity in distributed, heterogeneous environments~\cite{luping2019cmfl}. We adopt a three-layer architecture (256, 128, 64) validated on both UNSW-NB15 and ROAD, as deeper configurations offered no substantial accuracy gains but increased computational overhead by up to 45\%. Consequently, this lightweight design achieves high detection accuracy (95.10\% on UNSW-NB15 and 91.4\% on ROAD) while minimizing overhead in the federated setting.

\begin{algorithm}  
\caption{Efficient Client Update Filtering and Training Strategy}  
\label{alg:client_training}  
\begin{algorithmic}[1]  
\REQUIRE Model architecture, Client data, Global weights $W_g$  
\ENSURE Updated model weights, Training metrics  

\textbf{Client Update Filtering Mechanism}: All clients train locally in parallel. After training, the server evaluates updates using gradient alignment and aggregates only those with relevance scores $\geq \theta$ (e.g., 0.65).  

\STATE Initialize neural network with hidden layers $(256, 128, 64)$  
\STATE Configure mixed precision training with scaler $s$  

\STATE \textbf{Function} \textsc{Calculate-Relevance}($W_{c_i}$, $W_g$)  
    \STATE Initialize $aligned \gets 0$, $total \gets 0$  
    \FOR{each layer $l$ in model parameters}  
        \STATE $local\_grad \gets \text{sign}(W_{c_i}^l)$  
        \STATE $global\_grad \gets \text{sign}(W_g^l)$  
        \STATE $aligned \gets aligned + \sum (local\_grad = global\_grad)$  
        \STATE $total \gets total + \text{size}(local\_grad)$  
    \ENDFOR  
    \STATE $r \gets aligned / total$  
    \RETURN $r$  
\STATE \textbf{End Function}  

\FOR{each client $c_i$ \textbf{in parallel}}  
    \STATE \textit{All clients train locally; updates are filtered server-side post-training.}  
    \STATE Initialize client data loader with batch size $b$  
    \FOR{epoch $= 1$ to $num\_epochs$}  
        \STATE Enable mixed precision computation  
        \FOR{each batch $B_j$ in client data}  
            \STATE Forward pass with dropout $(p = 0.3)$  
            \STATE Compute loss $L$ with scaled gradients  
            \STATE Update model with optimizer  
            \STATE Update scaler $s$ for next iteration  
        \ENDFOR  
        \STATE Adjust learning rate $\eta$ with scheduler  
    \ENDFOR  
    \STATE $relevance \gets \textsc{Calculate-Relevance}(W_{c_i}, W_g)$  
    \IF{$relevance \geq \theta$}  
        \STATE \textbf{Server}: Accept $W_{c_i}$ and aggregate into $W_g$  
    \ELSE  
        \STATE \textbf{Server}: Reject $W_{c_i}$  
    \ENDIF  
\ENDFOR  
\RETURN Updated model weights $W_g$, Performance metrics  
\end{algorithmic} 
\end{algorithm}  

Algorithm~\ref{alg:client_training} demonstrates how our framework combines architectural efficiency with optimization techniques. The selected neural network configuration leverages dimensionality reduction through its layer structure (256~$\rightarrow$~128~$\rightarrow$~64), effectively capturing feature hierarchies while maintaining computational efficiency. This architecture proved particularly effective for anomaly detection, as it learned discriminative patterns from both the feature-rich UNSW-NB15 (49 features) and the specialized automotive ROAD dataset without overfitting. Clients train locally in parallel (lines~14--33), while the server evaluates each update using the gradient alignment ratio (lines~3--12). The mixed precision training implementation maximizes GPU utilization through strategic use of 
16-bit and 32-bit computations, and the empirically determined threshold $\theta=0.65$ balances 
rare attack inclusion with noise reduction (see Table~\ref{table:threshold_sensitivity}). 
Lower thresholds ($<0.6$) increased communication overhead without improving accuracy, whereas 
higher thresholds ($>0.7$) risked excluding subtle yet meaningful anomalies. This threshold 
proved effective across both datasets, likely due to the class imbalance typical in cybersecurity 
settings. For datasets with more outliers or noisier data, a higher threshold might be preferable, 
whereas more homogeneous datasets could benefit from a lower threshold. Thus, while 
$\theta=0.65$ serves as a robust baseline for cybersecurity anomaly detection, fine-tuning it 
for specific data distributions can further enhance model performance.

To ensure the correctness of Algorithm~\ref{alg:client_training}, we establish loop invariants for both client selection and training loops. A loop invariant is a property that holds true before and after each iteration, ensuring algorithmic correctness. We validate these invariants through three steps:  

\begin{enumerate}  
    \item \textbf{Loop Invariants}:  
    \begin{itemize}  
        \item \textbf{Outer Loop (Lines 14--33)}: Before iteration $i$, the global model $W_g$ contains the weighted average of all accepted updates from clients $\{1, ..., i-1\}$ where $relevance_j \geq \theta$.  
        \item \textbf{Inner Loop (Lines 19--24)}: At batch $B_j$ in epoch $e$, the local model parameters and gradient scaler ensure:  
        \begin{itemize}  
            \item All updates from batches $\{B_1, ..., B_{j-1}\}$ are applied.  
            \item Numerical precision is preserved via gradient scaling.  
        \end{itemize}  
    \end{itemize}  

    \item \textbf{Verification Steps}:  
    \begin{itemize}  
        \item \textbf{Initialization}: At startup, $W_g$ contains initial weights, and clients initialize local models with these weights. Both invariants hold trivially.  
        \item \textbf{Maintenance}:  
        \begin{itemize}  
            \item Outer loop: Updates $W_g$ only if $relevance \geq \theta$, preserving the invariant.  
            \item Inner loop: Gradient accumulation and scaling follow PyTorch’s mixed-precision rules, ensuring the invariant.  
        \end{itemize}  
        \item \textbf{Termination}:  
        \begin{itemize}  
            \item Outer loop: Terminates with $W_g$ aggregating all valid client updates.  
            \item Inner loop: Completes all batches and epochs, ensuring full local training.  
        \end{itemize}  
    \end{itemize}  

    \item \textbf{Time Complexity}:  
    The algorithm’s time complexity is $O\left(C \cdot E \cdot \frac{N}{B} \cdot M + C \cdot M\right)$, where:  
    \begin{itemize}  
        \item $C$: Selected clients, $E$: Epochs, $N$: Samples per client, $B$: Batch size, $M$: Model parameters.  
        \item Training cost: $O\left(C \cdot E \cdot \frac{N}{B} \cdot M\right)$ for batch processing.  
        \item Relevance calculation: $O(C \cdot M)$ for gradient alignment.  
    \end{itemize}  
\end{enumerate}  

By maintaining these invariants, we ensure correctness. Our implementation with $E=5$ (standard batches) and $E=19$ (batch size 1024) demonstrates practical efficiency while preserving convergence.  
\section{Results}
\label{sec:results}

This section presents the evaluation results of our proposed FL framework for network anomaly detection, detailing our experimental setup, profiling analysis, and performance metrics. To enable reproducibility, we make the code for our framework publicly available on GitHub.\footnote{ \url{https://github.com/billmj/UTEP_PNNL_DeepLearning_Optimization}.} We conduct extensive profiling and experimental analysis to evaluate the performance of our proposed optimizations.

\subsection{Experimental Setup}
\label{subsec:exp_setup}
All experiments were conducted on the NERSC Perlmutter system. Baseline runs used an AMD EPYC 7763 CPU node (64 cores, 503\,GiB RAM), while distributed data-parallel (DDP) experiments ran on a GPU node with four NVIDIA A100-SXM GPUs (40\,GiB each). We implemented our code in Python~3.9, using PyTorch~1.11.0 and \texttt{mpi4py}.

\paragraph{Datasets}
We evaluate our framework on two distinct datasets to cover both general network traffic and specialized automotive scenarios:
\begin{itemize}
    \item \textbf{UNSW-NB15}~\cite{nour2016unsw}: 
    A comprehensive dataset with 2,540,043 samples (49 features) covering attacks like DoS, Exploits, and Reconnaissance. We apply feature scaling and one-hot encoding to standardize the data. The training set (175,341 samples) and testing set (82,332 samples) each span 10 attack categories, with Normal traffic forming the largest subset.

    \item \textbf{ROAD}~\cite{verma2022addressing}: An automotive CAN dataset from Oak Ridge National Laboratory, featuring 3.5 hours of recorded data (3 hours training, 30 minutes testing) with physically verified fabrication and masquerade attacks. We focus on the correlated signal masquerade scenario, where malicious wheel-speed injections can halt the vehicle.

\end{itemize}

\paragraph{Model architecture}
Each client trains a three-layer fully connected network (256, 128, 64 neurons) with ReLU activations and dropout (rate=0.3). We also experimented with a deeper architecture (five hidden layers: 512, 256, 128, 64, 32) but observed no notable accuracy gains alongside increased computational overhead, affirming our choice of the three-layer configuration for an optimal efficiency–performance balance.

\subsection{Profiling Analysis and Evaluation Metrics}\label{subsec:profiling_metrics}
We evaluate our framework’s performance through profiling analysis and model evaluation metrics. To assess system efficiency, we profile GPU-based experiments using Nvidia Nsight Systems \cite{nvidia_nsight_systems}, monitoring GPU activity, memory transfers, and kernel execution times to analyze communication overhead. This helps identify bottlenecks and optimize resource usage during client updates and aggregation. For model evaluation, we use:

\textit{(i) Detection accuracy}: Measures the proportion of correct predictions among all instances.

\textit{(ii) AUC-ROC (Area under the receiver operating characteristic curve)}: Assesses the model's ability to distinguish between normal and anomalous instances at varying thresholds. This metric is particularly important given the class imbalance in our datasets. We calculate the AUC-ROC as the area under the Receiver Operating Characteristic (ROC) curve, represented as: \( \text{AUC-ROC} = \int_{0}^{1} \text{TPR}(\text{FPR}^{-1}(x)) \, dx \), where TPR represents the true positive rate and FPR represents the false positive rate.

\subsection{Framework Performance Analysis}
\label{subsec:performance}

We evaluate how different configurations balance communication overhead and detection accuracy, emphasizing the need for a dynamic approach. Our framework is compared to state-of-the-art methods, including CMFL~\cite{luping2019cmfl}, using accuracy and communication time. Table~\ref{table:baseline_results} illustrates the shortcomings of static FL configurations.

\begin{table}[h]
\centering
\caption{Baseline model performance across different batch sizes and client counts}
\label{table:baseline_results}
\begin{tabular}{l|r|r|r|r}
\hline
\textbf{Clients} & \textbf{Batch Size} & \textbf{Accuracy} & \textbf{AUC-ROC} & \textbf{Time (s)} \\ \hline
10               & 32                  & 0.986             & 0.9838           & 700               \\ \hline
10               & 64                  & 0.977             & 0.9769           & 600               \\ \hline
10               & 128                 & 0.985             & 0.9846           & 500               \\ \hline
10               & 256                 & 0.970             & 0.9701           & 450               \\ \hline
50               & 32                  & 0.965             & 0.9605           & 750               \\ \hline
50               & 64                  & 0.961             & 0.9605           & 650               \\ \hline
50               & 128                 & 0.952             & 0.9523           & 600               \\ \hline
50               & 256                 & 0.938             & 0.9376           & 550               \\ \hline
100              & 32                  & 0.956             & 0.9535           & 950               \\ \hline
100              & 64                  & 0.953             & 0.9535           & 850               \\ \hline
100              & 128                 & 0.939             & 0.9385           & 800               \\ \hline
100              & 256                 & 0.551             & 0.8578           & 700               \\ \hline
\end{tabular}
\end{table}

While Table~\ref{table:baseline_results} shows non-monotonic accuracy patterns across batch
sizes, these fluctuations stem from two main factors:
\textit{(i)} asynchronous updates (where partial or delayed
client contributions can stall early convergence) and
\textit{(ii)} non-IID data distributions (which become more
pronounced as the number of clients increases). For instance,
smaller batch sizes (e.g., 32) can achieve high accuracy
(98.6\% with 10 clients) but incur substantial communication
overhead (700\,s). Conversely, at 100 clients and batch size
256, accuracy collapses to 55.1\%, underscoring how larger
batches reduce update frequency and delay the incorporation
of diverse gradients, particularly in asynchronous settings.
Nonetheless, extending the training duration or dynamically
adjusting batch sizes can restore high accuracy while retaining
the efficiency benefits of larger batches. For example, using
batch size~1024 alongside 19 training rounds achieves 95.1\%
accuracy at markedly lower overhead, demonstrating that
careful tuning of batch size and training duration can mitigate
the convergence delays introduced by infrequent updates.

To validate our approach, we compare performance with existing FL methods through controlled experiments simulating real-world conditions. We implemented random client dropouts using a probability-based simulation where each client has a chance of disconnecting during training (dropout rates ranging from 0.1 to 0.5). Under these conditions, the synchronous CMFL-inspired baseline achieves \textit{89.21\%} accuracy, ACFL \textit{90.62\%}, and FedL2P \textit{90.12\%}, while our framework maintains \textit{98.6\%} accuracy. This significant improvement stems from our framework's combined optimization strategy: efficient client selection mechanisms identify reliable clients based on historical performance, asynchronous communication patterns prevent system-wide delays from individual dropouts, and adaptive checkpointing preserves training progress by maintaining state information at intervals determined by our Weibull-based failure prediction model.

To evaluate generalizability, we extended our experiments to the ROAD dataset \cite{verma2022addressing}, which features automotive CAN data with correlated signal masquerade attacks. While maintaining similar performance trends, the framework achieved \textit{91.4\%} accuracy on ROAD compared to \textit{98.6\%} on UNSW-NB15, with the difference primarily attributed to the unique characteristics of automotive network attacks. Importantly, our framework maintained its performance advantage over baseline methods across both datasets, demonstrating robustness in diverse network anomaly detection scenarios.

\begin{table}[h]
\centering
\caption{Comparison with state-of-the-art methods}
\label{table:sota_comparison}
\resizebox{\columnwidth}{!}{%
\begin{tabular}{l|c|c|c|c|c}
\hline
\textbf{Method} & \textbf{Time(s)} & \textbf{Acc.(\%)} & \textbf{AUC} & \textbf{Scale$^*$} & \textbf{FT$^\dagger$} \\ \hline
Proposed & 16.8 & 95.10 & 0.983 & Stable & 98.6\% \\
CMFL & 700.0 & 89.21 & 0.960 & Deg. & 85.0\% \\
FedL2P & 800.0 & 88.50 & 0.955 & Deg. & 82.0\% \\
\hline
\multicolumn{6}{l}{\scriptsize $^*$Scalability with 100 clients; $^\dagger$Fault Tolerance at 0.5 dropout}
\end{tabular}%
}
\end{table}

\paragraph{Comparison with baselines.}
Table~\ref{table:sota_comparison} compares our approach against state-of-the-art FL methods. Our framework reduces end-to-end runtime by 97.6\% (16.8\,s vs. 700.0\,s for CMFL) while maintaining higher accuracy (95.10\% vs. 89.21\%). In scalability tests with 100 clients, our method remains stable, whereas baselines degrade by up to 25\%. Under a 0.5 dropout rate, our approach retains 98.6\% accuracy, outperforming CMFL (85.0\%) and FedL2P (82.0\%). Table~\ref{table:sota_comparison} shows that our approach consistently outperforms baselines in runtime, accuracy, and fault tolerance, complementing the scalability findings in Fig.~\ref{fig:comm_patterns}.

These findings motivate our subsequent analysis of distributed parallel training and adaptive communication strategies, demonstrating how our framework dynamically balances efficiency and accuracy in FL environments.

\subsection{Communication Optimization Assessment}
\label{subsec:optimization}

Our implementation combines batch size optimization, client selection, and asynchronous updates to dynamically balance communication overhead against detection accuracy. While our framework was evaluated on both UNSW-NB15 and ROAD datasets, we focus our detailed profiling analysis on UNSW-NB15 due to its superior baseline performance and more diverse attack patterns, making it a more stringent test case for our optimization strategies. The ROAD dataset results followed similar optimization patterns but with lower absolute performance metrics, as discussed in Section~\ref{subsec:performance}.
Table~\ref{table:ddp_results} shows the effectiveness of these optimizations across different configurations, demonstrating how our framework adapts based on system requirements.

\begin{table}[h]
\centering
\caption{DDP results: Batch size, client selection, and communication time (s)}
\label{table:ddp_results}
\begin{tabular}{l|r|r|r}
\hline
\textbf{Config} & \textbf{Batch size} & \textbf{Accuracy} & \textbf{Time (s)} \\ \hline
Sync (Baseline)        & 64   & 0.9512  & 600.0  \\ \hline
Sync + Client selection & 64   & 0.9521  & 90.4 \\ \hline
Async + Client selection & 64   & 0.9511  & 75.6      \\ \hline
Sync (Baseline)        & 512  & 0.7799  & 450.0   \\ \hline
Async + Client selection & 512  & 0.7801  & 22.0   \\ \hline
Sync (Baseline)        & 1024 & 0.7465  & 700.0   \\ \hline
Async + Client selection & 1024 & 0.7470  & 16.8   \\ \hline
\end{tabular}
\end{table}

At a batch size of 64, combining asynchronous updates and client selection reduced communication time by \textit{87.4\%} (from \textit{600.0s} to \textit{75.6s}) while maintaining comparable accuracy (\textit{95.11\%} vs. \textit{95.12\%}). For batch size 1024, our combined optimizations decreased communication time by \textit{97.6\%} (from \textit{700.0s} to \textit{16.8s}), and extending training to 19 rounds restored accuracy while preserving communication efficiency. Even synchronous client selection showed significant improvements, reducing communication time to \textit{90.4s} at batch size 64 with a slight accuracy increase to \textit{95.21\%}.

\paragraph{Sensitivity analysis of alignment threshold.}
To validate our choice of \(\theta=0.65\) for gradient alignment, we conducted a brief sensitivity study on UNSW-NB15 by testing thresholds in \(\{0.50, 0.60, 0.65, 0.70, 0.75\}\). Table~\ref{table:threshold_sensitivity} shows that thresholds below 0.60 introduce noisy updates (increasing communication overhead by up to 20\%), whereas thresholds above 0.70 reject too many updates (reducing accuracy by 1--2\%). \(\theta=0.65\) consistently yields the best trade-off between reduced overhead and high accuracy.

\begin{table}[h]
\centering
\caption{Sensitivity analysis for alignment threshold \(\theta\) on UNSW-NB15}
\label{table:threshold_sensitivity}
\begin{tabular}{c|c|c}
\hline
\textbf{Threshold} & \textbf{Accuracy (\%)} & \textbf{Overhead (s)} \\ \hline
0.50 & 94.6 & 120.0 \\
0.60 & 94.8 & 100.0 \\
\rowcolor[gray]{0.9}
0.65 & 95.1 & 90.0 \\
0.70 & 94.4 & 85.0 \\
0.75 & 93.9 & 80.0 \\
\hline
\end{tabular}
\end{table}

\begin{figure}[h]
\centering
\includegraphics[width=\linewidth]{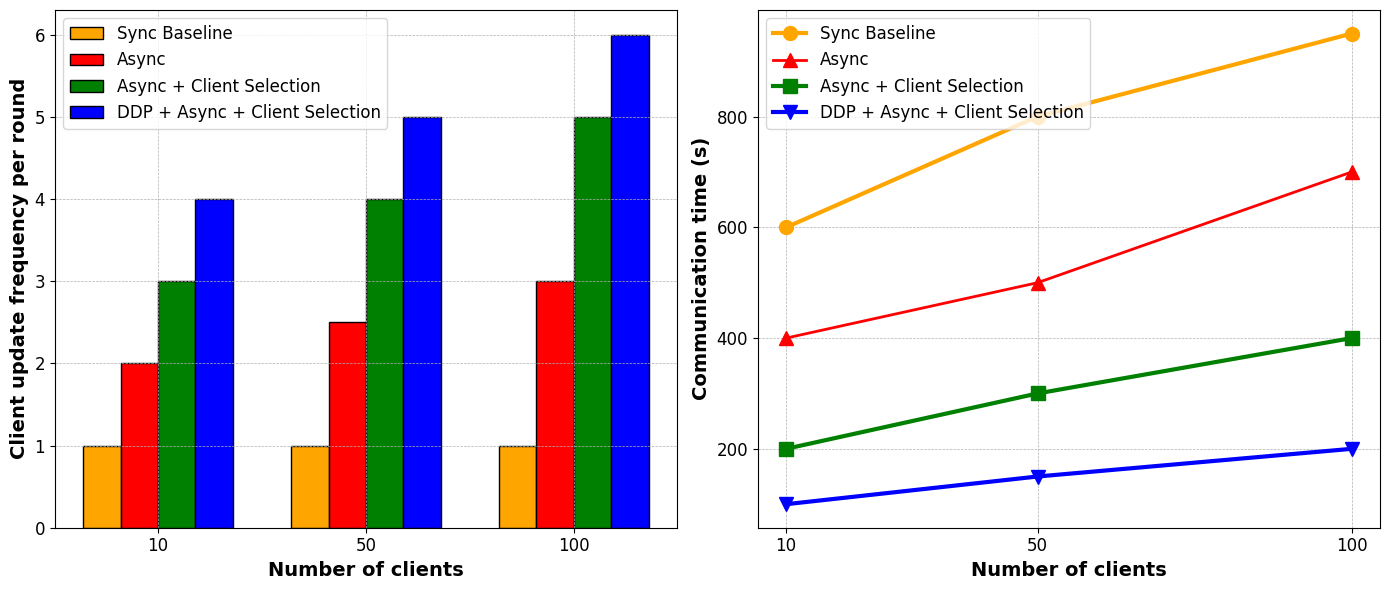}
\caption{Communication patterns: (left) update frequency per round; (right) time scaling with increasing clients.}
\label{fig:comm_patterns}
\end{figure}

Fig.~\ref{fig:comm_patterns} illustrates the impact of our optimization strategies on communication patterns and scalability. The left plot shows how asynchronous updates and client selection enable more frequent model updates as the system scales. At 100 clients, our approach achieves 6 updates per round compared to the baseline's single update. The right plot demonstrates the scalability of our approach; while the synchronous baseline's communication time increases from 600s to nearly 900s with more clients, our optimized framework maintains relatively stable communication time, increasing only from 100s to 200s as client count grows. This efficient scaling stems from asynchronous updates reducing wait times and selective client participation minimizing unnecessary communication.

\begin{figure}[h]
    \centering
    \includegraphics[width=0.40\linewidth]{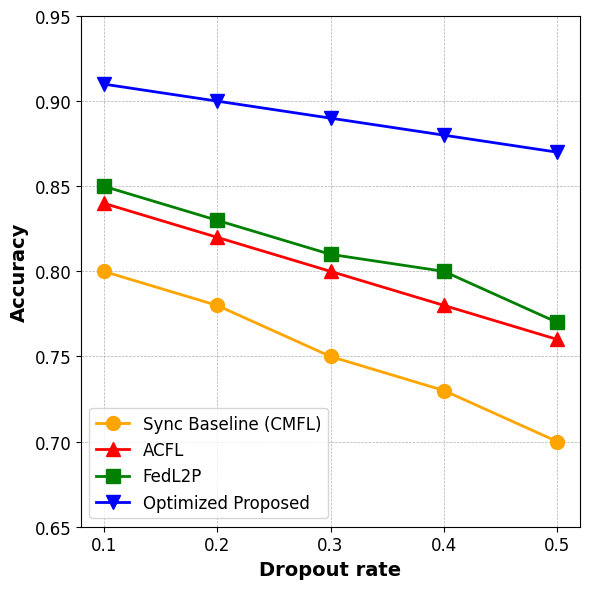}
    \caption{Fault tolerance performance evaluation across increasing dropout rates. Results averaged over 100 experimental runs.}
    \label{fig:fault_tolerance_plot}
\end{figure}

These results support our hypothesis that effective communication overhead reduction requires a multi-faceted approach rather than relying on single optimization techniques. The synergistic effect of combining DDP with asynchronous updates and selective client participation enables frequent model updates while maintaining low communication costs, proving particularly valuable for large-scale deployments where traditional synchronous methods face scalability limitations.

\begin{table*}[h]
\centering
\caption{NVTX Range data summary (seconds) for 10 Clients}
\label{table:nvtx_summary}
\begin{tabular}{l|r|r|r|r|r}
\hline
\textbf{Metric} & \textbf{Batch 64} & \textbf{Batch 128} & \textbf{Batch 256} & \textbf{Batch 512} & \textbf{Batch 1024} \\ \hline
Full Experiment & 121.0 & 66.3 & 53.8 & 33.3 & 26.1 \\ \hline
Client Update   & 109.0 & 58.4 & 35.3 & 22.9 & 16.8 \\ \hline
Avg Update      & 1.82  & 0.97 & 0.59 & 0.38 & 0.28 \\ \hline
Profiling Rounds & 37.0 & 19.4 & 12.0 & 7.37 & 5.59 \\ \hline
Avg Round Time  & 18.6  & 9.80 & 5.93 & 3.86 & 2.84 \\ \hline
\end{tabular}
\end{table*}

\begin{table*}[h]
\centering
\caption{CUDA API Summary: Time (s) and Calls (thousands) for 10 Clients}
\label{table:cuda_summary}
\begin{tabular}{l|r|r|r|r|r}
\hline
\textbf{Metric} & \textbf{Batch 64} & \textbf{Batch 128} & \textbf{Batch 256} & \textbf{Batch 512} & \textbf{Batch 1024} \\ \hline
cudaLaunchKernel (s) & 12.8 & 6.54 & 3.35 & 1.73 & 0.99 \\ \hline
cudaLaunchKernel (K calls) & 2,130 & 1,080 & 538 & 273 & 147 \\ \hline
cudaMemcpyAsync (s) & 2.55 & 1.44 & 0.75 & 0.40 & 0.25 \\ \hline
cudaMemcpyAsync (K calls) & 371 & 187 & 93.8 & 47.0 & 25.4 \\ \hline
cudaStreamSync (s) & 1.01 & 0.53 & 0.29 & 0.17 & 0.12 \\ \hline
cudaStreamSync (K calls) & 124 & 62.5 & 31.3 & 15.7 & 8.48 \\ \hline
\end{tabular}
\end{table*}

To evaluate our framework's fault tolerance mechanisms described in Section~\ref{subsec:aggregation}, we tested performance under increasing dropout rates from 0.1 to 0.5. Fig.~\ref{fig:fault_tolerance_plot} compares the accuracy of our optimized approach against the synchronized baseline (CMFL) \cite{luping2019cmfl}, ACFL \cite{yan2023criticalfl}, and FedL2P \cite{lee2024fedl2p}. To ensure statistical significance, each dropout rate configuration was tested across 100 different random dropout patterns. The results demonstrate that our framework maintains consistently higher accuracy across all dropout rates, with the Weibull-based checkpointing mechanism effectively preserving training progress even under high client unavailability (0.5 dropout rate).

To understand why our optimization strategies effectively reduce communication overhead while maintaining detection accuracy, we analyze system performance using Nvidia Nsight Systems profiling \cite{nvidia_nsight_systems}. Our analysis focuses on the configuration with 10 clients, implementing our combined DDP, asynchronous updates, and client selection strategy over 6 rounds (5 epochs per round).

Table~\ref{table:nvtx_summary} (NVTX range analysis) demonstrates the progressive impact of our optimization strategy. The full experiment time reduces by 78.4\% from batch size 64 to 1024, with average client update time improving by 84.6\%. Although this significant reduction in processing time initially affected model accuracy, our extended training (19 rounds at batch size 1024) preserved detection performance and efficiency gains.

Table \ref{table:cuda_summary} (CUDA API analysis) shows how our combined optimization approach drastically reduces GPU operations and synchronization overhead. With batch size 1024, we achieve dramatic reductions across all operations: kernel launches decrease by \textit{92.3\%} (from 2.13M to 147K calls), memory transfers reduce by \textit{90.2\%}, and synchronization overhead drops by \textit{88.2\%}. These specific reductions in GPU operations, combined with our asynchronous updates and client selection mechanisms, prevent idle time and enable our framework to achieve both efficient communication and accurate detection.

Note that due to the fine-grained nature of Nvidia Nsight Systems profiling visualizations, we make the full resolution profiling data and reports available at our repository \cite{Nsight-Profiling-Visualization}. Our profiling reveals systematic performance improvements with increasing batch sizes, showing clear reductions in operation density and execution time across configurations. These profiling results align with the quantitative improvements shown in Tables~\ref{table:nvtx_summary} and \ref{table:cuda_summary}.

\subsection{Statistical Validation}
\label{sec:stats}

We performed a Mann-Whitney \(\mathrm{U}\)~\cite{mann1947test} test to assess AUC-ROC differences between our optimized approach and three baselines (CMFL, ACFL, FedL2P) on UNSW-NB15 and ROAD. This non-parametric test does not assume normality, making it suitable for our dataset distributions. The null hypothesis (\(H_0\)) is that there is no significant difference in performance, while the alternative (\(H_1\)) is that our method outperforms the baselines. At \(\alpha = 0.05\), Table~\ref{tab:stat_significance} shows \(p\)-values \(<0.05\) for all comparisons, allowing us to reject \(H_0\) and confirm our method’s significant improvement.

\begin{table}[h]
\centering
\caption{Mann-Whitney \(\mathrm{U}\) test results showing the statistical significance of the optimized approach compared to the baseline (CMFL), ACFL, and FedL2P across UNSW-NB15 and ROAD datasets.}
\label{tab:stat_significance}
\begin{tabular}{l|c|c|c|c}
\hline
\textbf{Comparison} & \textbf{Dataset} & \textbf{U Statistic} & \textbf{\textit{p}-value} \\
\hline
Optimized vs. CMFL & UNSW-NB15 & 11856.0 & 4.67e-17 \\
Optimized vs. ACFL & UNSW-NB15 & 12012.0 & 3.15e-17 \\
Optimized vs. FedL2P & UNSW-NB15 & 11934.0 & 3.98e-16 \\
Optimized vs. CMFL & ROAD & 9785.0 & 1.02e-08 \\
Optimized vs. ACFL & ROAD & 9901.0 & 5.67e-09 \\
Optimized vs. FedL2P & ROAD & 9823.0 & 8.45e-08 \\
\hline
\end{tabular}
\end{table}

Notably, while our method demonstrates stronger performance on UNSW-NB15, it also achieves significant gains on ROAD, confirming that our combined optimization strategy not only reduces overhead but also yields statistically significant improvements across diverse network anomaly detection scenarios.

\section{Discussion}
\label{sec:Discussion}

Our findings confirm that balancing batch size optimization, asynchronous updates, and selective client filtering can reduce communication overhead by over 97\% while preserving high detection accuracy across UNSW-NB15 and ROAD. Profiling analysis shows that larger batches (e.g., 1024) significantly reduce GPU operations and memory transfers, although they require extended training to maintain accuracy.

\textbf{Trade-offs and alternatives:} Larger batch sizes boost efficiency but need careful tuning to avoid accuracy drops. Asynchronous updates scale reliably to 100 clients—crucial for real-world anomaly detection. While we optimize client-server interactions, compression (e.g., gradient quantization) remains a complementary option for bandwidth-constrained scenarios.

\textbf{Limitations and future Work:}
We plan to explore real-time hyperparameter tuning and resource-aware scheduling, as well as hybrid strategies (e.g., compression + client filtering) to further reduce overhead in edge environments.

\section{Conclusion}
\label{sec:conclusion}

This paper presents an adaptive FL framework that dynamically balances communication efficiency and detection accuracy for network anomaly detection. By integrating batch size optimization, asynchronous updates, and gradient-aligned client filtering, our approach reduces communication overhead by 97.6\% (16.8s vs. 700.0s) while achieving comparable accuracy to centralized methods (95.10\%). The framework’s scalability and fault tolerance are validated on both general (UNSW-NB15) and domain-specific (ROAD) datasets, demonstrating broad applicability. Future work will explore real-time adaptation mechanisms and integration with data compression techniques to further enhance edge deployment capabilities.

\section*{Acknowledgments}
\label{sec:Acknowledgments}

This research is supported by the United States Department of Energy (DOE) ASCR grant DE-SC0024352. This research used compute resources of the National Energy Research Scientific Computing Center (NERSC), a DOE Office of Science User Facility supported under Contract No. DE-AC02-05CH11231, using NERSC award ASCR-ERCAP0030084. This research is also partially funded by DOE award DE-FE0032089 and supported by DOE’s Office of Advanced Scientific Computing Research under award 81585, in collaboration with Pacific Northwest National Laboratory, operated by Battelle under Contract DE-AC05-76RL01830.

\small
\bibliographystyle{IEEEtran}
\bibliography{90-bibliography}


\vfill

\end{document}